\documentclass[aps,prd,a4paper,twocolumn,amsmath,showpacs,superscriptaddress,nofootinbib,preprintnumbers]{revtex4-1}

\usepackage{graphicx,ulem}
\usepackage{longtable}
\usepackage{float}
\usepackage{dcolumn}
\usepackage{graphics,epsfig}
\usepackage{amsmath,amssymb,latexsym,mathrsfs}
\usepackage{bm}
\usepackage{color}

\newcommand{\MeV}{\,\mathrm{MeV}}
\newcommand{\eV}{\,\mathrm{eV}}
\newcommand{\iso}{\mathcal{S}}
\newcommand{\reff}[1]{(\ref{#1})}
\newcommand{\Neff}{N_\mathrm{eff}}
\newcommand{\DNeff}{\Delta N_\mathrm{eff}}
\def\he4{$^4$He}
\def\h2{$^2$H}

\begin{document}

\title{Future constraints on neutrino isocurvature perturbations in the curvaton scenario}


\author{Eleonora Di Valentino}
\affiliation{Physics Department and INFN, Universit\`a di Roma 
	``La Sapienza'', Ple.\ Aldo Moro 2, 00185, Rome, Italy}
\author{Massimiliano Lattanzi}
\affiliation{Dipartimento di Fisica G. Occhialini, Universit\`a Milano-Bicocca and INFN, sezione di Milano-Bicocca, Piazza della Scienza 3,
I-20126 Milano, Italy}
\author{Gianpiero Mangano}
\affiliation{INFN, Sezione di Napoli, Complesso Univ. Monte S. Angelo, Via Cintia, I-80126 Napoli, Italy}
\author{Alessandro Melchiorri}
\affiliation{Physics Department and INFN, Universit\`a di Roma 
	``La Sapienza'', Ple.\ Aldo Moro 2, 00185, Rome, Italy}
\author{Pasquale Serpico}
\affiliation{LAPTh, Univ. de Savoie, CNRS, B.P.110, Annecy-le-Vieux F-74941, France}

\date{\today}

\preprint{LAPTH-044/11}

\begin{abstract}
In the curvaton scenario, residual isocurvature perturbations can be imprinted in the cosmic neutrino component after the decay of the curvaton field, implying in turn a non-zero chemical potential in the neutrino distribution. We study the constraints that future experiments like Planck, SPIDER or CMBPol will be able to put on the amplitude of isocurvature perturbations in the neutrino component. We express our results in terms of the square root $\gamma$ of the non-adiabaticity parameter $\alpha$ and of the extra relativistic degrees of freedom $\DNeff$. Assuming a fiducial model with purely adiabatic fluctuations, we find that Planck (SPIDER) will be able to put the following upper limits at the 1$\sigma$ level: $\gamma \le 5.3\times 10^{-3}\,(1.2\times 10^{-2})$ and $\DNeff\le 0.16\,(0.40)$. CMBPol will further improve these constraints to $\gamma \le 1.5\times 10^{-3}$ and $\DNeff \le 0.043$.  Finally, we recast these bounds in terms of the background neutrino degeneracy parameter $\bar \xi$ and the corresponding perturbation amplitude $\sigma_\xi$, and compare with the bounds on $\bar \xi$ that can be derived from Big Bang Nucleosynthesis.
\end{abstract}

\maketitle

\section{Introduction}\label{intro}
In single-field inflationary models, the same field is responsible for driving an accelerated expansion stage
and for the generation of a nearly scale invariant primordial perturbation spectrum. As there is only one degree of freedom, this class of models predicts that 
perturbations are necessarily adiabatic, i.e., the ratio between the number densities of the different particle species is spatially homogeneous. Significant non-Gaussianities in the fluctuations are also excluded. To date, both these features, adiabaticity and gaussianity, are consistent with data.
 
However, the presence of a significant, albeit sub-dominant, non-adiabatic (otherwise called \textit{isocurvature}) perturbation component cannot be excluded, see e.g. \cite{Crotty:2003rz,Beltran:2004uv,Moodley:2004nz,Bean:2006qz,Trotta:2006ww,Komatsu:2010fb}. This component must be necessarily related to some extra field other than the inflaton, as in multifield inflationary models, where non trivial trajectories in field space are possible. Since in these cases the adiabatic and isocurvature fluctuations
would be related to different fields, generating a sizeable isocurvature fluctuation requires in general a certain amount of fine-tuning.

A different mechanism for isocurvature modes production has been proposed in \cite{Lyth:2001nq,Lyth:2002my}. While the inflaton is only responsible for driving the exponential expansion, primordial fluctuations are generated by a ``curvaton'' field. The initial isocurvature perturbation in the curvaton is then converted into an adiabatic component after  inflaton decay. This model allows  for some residual isocurvature components imprinted in the other components of the cosmological fluid, cold dark matter, baryons and neutrinos, after curvaton decay. In particular neutrino isocurvature perturbations requires a non vanishing chemical potential for their background distribution in phase space. Probing their non adiabatic perturbations is thus, a way to constrain the lepton number in neutrino sector. An analysis of the bounds on neutrino isocurvature perturbations using recent data is presented in Ref. \cite{Gordon:2003hw}, while limits on isocurvature perturbations in an extra radiation component (not necessarily related to neutrinos) have been derived in Ref. \cite{Kawasaki:2011rc}. Quite interestingly, a future detection of isocurvature perturbations will allow for a  reconstruction of the inflationary potential \cite{Easson:2010uw}.

The aim of the present paper is to assess the capability of future cosmic microwave background (CMB) experiments like Planck  \cite{:2006uk, Tauber:2010aa}, SPIDER \cite{Fraisse:2011xz} and CMBPol \cite{Bock:2009xw} to constrain simultaneously the amplitude of isocurvature perturbations in the neutrino component and the extra energy density associated to the neutrino chemical potential. The bounds can then be translated into constraints on the neutrino chemical potential to temperature ratio $\xi_i$ ($i={e,\mu,\tau}$) and the corresponding perturbation amplitudes. These are  complementary to bounds on the $\xi_i$'s which can be derived using Big Bang Nucleosynthesis (BBN). Light nuclei yields in fact, are quite strongly influenced by neutrino asymmetries, mainly in the $\nu_e$ sector, see e.g. \cite{Iocco:2008va}. Since flavour oscillations are efficient in mixing different flavour distributions, the three parameters are driven to almost the same value at the onset of BBN \cite{Dolgov:2002ab}, with possible differences in the $e$ and $\mu,\tau$ sectors which cannot be larger than few percents depending on the value of the $\theta_{13}$ mixing angle \cite{Mangano:2010ei,Mangano:2011ip}.

The paper is organized as follows. In Section \ref{nu} we review the neutrino isocurvature perturbations which are generated in the curvaton scenario. 
Section \ref{result} contains a forecast analysis of bounds on these perturbations from future experiments, while a comparison with the corresponding BBN constraints is described in Section \ref{BBN}. Our conclusions are reported in Sec.~\ref{Conclusions}.

\section{Neutrino isocurvature perturbations}\label{nu}

Density perturbations are conveniently described in terms of the gauge-invariant quantity $\zeta$   \cite{Bardeen:1983qw,Mukhanov:1990me,Wands:2000dp}
\begin{equation}
\zeta = -\psi - H \frac{\delta\rho}{\dot\rho} \, ,
\end{equation}
where $\psi$ is the (gauge-dependent) curvature perturbation, $H$ the Hubble parameter, 
$\rho$ the total energy density, and the dot denotes derivatives with respect to the cosmological time $t$.

The quantity $\zeta$ describes the curvature perturbation on slices of uniform total density. In the case of multicomponent
fluids, it is useful to define quantities $\zeta_i$ describing the curvature perturbation on slices of uniform density of the $i$-th 
component
\begin{equation}
\zeta_i = -\psi - H\frac{\delta\rho_i}{\dot\rho_i} \, .
\end{equation}¥

An adiabatic fluctuation is defined as one for which the ratios $\delta\rho_i/\dot \rho_i$ are all the same, so that $\zeta_i = \zeta$ for all components. Correspondingly, a nonadiabatic (or isocurvature) fluctuation $\iso_i$ in the $i$-th fluid component is defined as the relative entropy fluctuation with respect to photons:
\begin{equation}
\iso_i\equiv 3(\zeta_i - \zeta_\gamma)\,.
\end{equation}¥

In the following, we shall consider neutrinos with an equilibrium distribution function
\begin{equation}
f_i(E) = \left[\exp(E/T_\nu\mp\xi_i)\right]^{-1} \, \label{eq:FD}\,,
\end{equation}
where $T_\nu$ is their temperature, and $\xi_i=\mu_i/T_\nu$, $\mu_i$ being the chemical potential. The index $i$ runs over the three standard model neutrino families, $i=e,\,\mu,\,\tau$, and the minus (plus) sign is for neutrinos (antineutrinos). Notice that the existence of neutrino isocurvature perturbations necessarily implies a non zero lepton asymmetry in the neutrino sector,  $n_L\equiv n_\nu - n_{\bar \nu}$, unless the asymmetries in the three flavours exactly cancel. At this stage, we have allowed for the possibility of the three neutrino families having different chemical potentials. The neutrino temperature is $T_\nu = T_\gamma$ until the time of electron-positron annihilation, occurring at $T_\gamma \simeq 1\,\MeV$ (shortly after neutrino decoupling), while at later times it is given by $T_\nu = (4/11)^{1/3} T_\gamma$, up to tiny corrections  due to neutrino reheating at the $e^\pm$ annihilation stage~\cite{Mangano:2005cc}.

Given the distribution function Eq. \reff{eq:FD}, the energy density $\rho_i\equiv \rho_{\nu_i}+\rho_{\bar\nu_i}$ in the 
high-temperature limit $T_\nu\gg m_\nu$ writes \cite{Lesgourgues:1999wu}:
\begin{align}
&\rho_i = \frac{7\pi^2}{120} A_i \, T_\nu^4=\frac{7}{8}A_i \left(\frac{T_\nu}{T_\gamma}\right)^4 \rho_\gamma \; , 
\end{align}¥
where
\begin{align}
&A_i \equiv \left[1+\frac{30}{7}\left(\frac{\xi_i}{\pi}\right)^2+\frac{15}{7} \left(\frac{\xi_i}{\pi}\right)^4 \right] \; , \label{eq:Ai}
\end{align}¥

When dealing with cosmological neutrinos, it is customary to define the effective number of neutrino families $\Neff$ as the ratio between
the total neutrino density and the density of a single non-degenerate ($\xi=0$) neutrino species in thermal equilibrium at $T_\nu = (4/11)^{1/3} T_\gamma$.
In the standard cosmological scenario $\Neff=3.046$, see \cite{Mangano:2005cc}, and any deviation $\DNeff$ from this value indicates the presence of an extra energy density of relativistic particles in the early Universe. It is clear, from our definition, that  $\Neff = \sum_{i} A_i$.
We can thus relate the isocurvature perturbation in the total neutrino density to the fluctuations $\delta\Neff^{(i)}$:
\begin{equation}
\iso_\nu=3(\zeta_\nu - \zeta_\gamma) \simeq \frac{\sum_i\delta\Neff^{(i)}}{4 \Neff}\,.\label{eq:isonu}
\end{equation}
\section{CMB constraints and forecast}\label{result}
In the following, lacking a better theoretical motivation, for simplicity we shall assume that both the average values and the fluctuations in the chemical potentials are flavor blind, i.e. $\bar \xi_e = \bar \xi_\mu = \bar\xi_\tau = \bar\xi$, and similarly for the $\delta \xi$'s. Also, we assume that  fluctuations in the neutrino degeneracy parameter are gaussian distributed with variance $\sigma^2_\xi$ around the mean $\bar\xi$. In general, both quantities can have a scale and epoch dependence. 

Conventionally, rather than in terms of $\iso_\nu$ of Eq.~\reff{eq:isonu}, in CMB studies the ``non-adiabaticity'' of perturbations is expressed in terms of the ratio of the power spectrum $P_\iso(k)$ of isocurvature perturbations to the curvature perturbation spectrum $P_\mathcal{\zeta}(k)$, evaluated at a fixed pivot wave number $k_0 = 0.002\,\mathrm{Mpc}^{-1}$. In particular, one introduces the quantity $\alpha$ defined by \cite{Bean:2006qz,Komatsu:2010fb}
\begin{equation}
\frac{\alpha(k_0)}{1-\alpha(k_0)} \equiv \frac{P_\iso(k_0)}{P_\mathcal{\zeta}(k_0)} \, ,\label{alphadef}
\end{equation}

Another necessary ingredient to be taken into account is the correlation between the adiabatic and isocurvature modes \cite{Langlois:1999dw,Langlois:2000ar,Gordon:2000hv}. Given the cross-correlation power spectrum $P_{\zeta\iso}(k)$, this is parameterized in terms of the cross-correlation coefficient $\beta$, defined as
\begin{equation}
\beta = \frac{P_{\zeta\iso}(k_0)}{\sqrt{P_\iso(k_0)P_\zeta(k_0)}} \, .
\end{equation}
We remark that we choose the sign convention for the curvature perturbation such that the temperature fluctuation at large scales is given by $\Delta T/T = \zeta/5 - 2\iso/5$. In terms of the variables used in the WMAP analysis \cite{Komatsu:2008hk,Komatsu:2010fb}, $\zeta=\tilde{\mathcal{R}}=-\mathcal{R}$, and our definition of $\beta$ coincides with the one used there. In this case, the physically observable effect is that correlated perturbations ($\beta>0$) reduce the temperature power spectrum at low multipoles. 

Given the above convention, the adiabatic and isocurvature fluctuations in the curvaton scenario are totally anticorrelated \cite{Lyth:2003ip,Moroi:2001ct,Bartolo:2002vf}, so that in the following we will always take $\beta=-1$. We also take the two power spectra to have the the same spectral tilt $n_s$:
\begin{equation}
\Delta^2_{\mathcal R,S}(k) \equiv \frac{k^3 P_{\mathcal R,S}}{2\pi^2}\propto k^{n_s-1} \, .
\end{equation}

Note that the CMB is sensitive to the parameters of the scenario not only via $\alpha$, but also via the total $\DNeff$ induced by the average value $ \bar\xi$ and, if sufficiently large, in principle also by the variance $\sigma^2_\xi$. For analyses or forecasts, one should thus consider the constraints in the $\DNeff-\alpha$ plane.

Current WMAP7 bounds on totally anticorrelated isocurvature perturbations are at the level $\alpha(k_0)<1.1\times 10^{-2}$~ (at the 95\% confidence level); inclusion of additional datasets can improve this bound by a factor 2 or 3 \cite{LAMBDA,Komatsu:2010fb}. For what concerns the effective number of relativistic species, WMAP7 observations only provide a lower limit $\Neff>2.7$; interestingly, when other cosmological measurements are considered, the result is $\Neff=4.34\pm0.9$ \cite{Komatsu:2010fb}, indicating that  the data seem to prefer $\DNeff>0$. 

The total CMB power spectrum can be parameterized in terms of the adiabatic, neutrino isocurvature density
 and totally anticorrelated spectra as follows
\begin{eqnarray}
 C_\ell&=& (1-\alpha)C_\ell^{ad} + \alpha C_\ell^{nid}+\nonumber\\
&&-2\sqrt{\alpha(1-\alpha)}C_\ell^{corr} \, ,
\label{eqn_cl}
\end{eqnarray}
with $\alpha$  defined in Eq. \reff{alphadef}. A shortcoming of this parameterization is that the partial derivative $\frac{\partial C_\ell}{\partial\alpha}$, needed for the Fisher matrix computation (see below), diverges for $\alpha=0$. This prevents the use of the Fisher matrix formalism for the fiducial value $\alpha=0$.  For this reason, we find convenient to introduce the auxiliary parameter $ \gamma=\sqrt{\alpha}$ and write
\begin{eqnarray}
 C_\ell&=& (1-\gamma^{2})C_\ell^{ad} + \gamma^{2} C_\ell^{nid}+\nonumber\\
&&-2 \gamma \sqrt{(1-\gamma^{2})}C_\ell^{corr} \, .
\label{eqn_cl}
\end{eqnarray}
It is straightforward to check that the partial derivative $\frac{\partial C_\ell}{\partial\gamma}$ is finite for $\gamma=0$.

In the following we will derive forecasts for the Planck \cite{:2006uk}, SPIDER \cite{Fraisse:2011xz} and the CMBPol \cite{Bock:2009xw} experiments.
The Planck  satellite \cite{:2006uk, Tauber:2010aa}, launched in May 2009, is currently measuring the CMB temperature and polarization fluctuations with unprecedented precision ($\Delta T/T \sim 2 \times 10^{-6}$) over the whole sky and down to very small angular scales ($\sim$5'). Planck measurements, planned to be publicly released to the scientific community in January 2013, will significantly improve the determination of cosmological parameters and will allow to test further the $\Lambda$CDM paradigm.  SPIDER \cite{Fraisse:2011xz}, scheduled to flight in 2012, is a ballon-borne polarimeter design to accurately measure the $B$-mode of CMB polarization down to $\ell\sim 100$.  Finally, CMBPol \cite{Bock:2009xw} is a next-generation satellite currently in the concept study phase.

In order to derive forecasts for these experiments, we use a Fisher matrix formalism, 
for three frequency channels for each experiment (the experimental specifications are listed in Table~\ref{tabexp}). 
\begin{table}[!htb]
\begin{center}
\begin{tabular}{rcccc}
Experiment & Channel[GHz] & FWHM & $\sigma_T [\mu K]$ &  $\sigma_P [\mu K]$\\
\hline
Planck & 217 & 5.0'& 13.1 & 18.5\\
$f_{sky}=0.65$
       & 143 & 7.0'& 5.99 & 8.48\\
       & 100 & 9.5'& 6.75 & 9.55\\
\hline
SPIDER 		& 280 & 17'&0.20  &0.29  \\
$f_{sky}=0.1$     & 150 & 30'& 0.08 & 0.11  \\
       			& 90 & 49'& 0.08  & 0.11 \\
\hline
CMBPol & 220 & 3.8' & 0.66 & 0.93\\
$f_{sky}=0.65$ 
       & 150 & 5.6' & 0.25 & 0.35\\
       & 100 & 8.4' & 0.22 & 0.31\\
\end{tabular}
\caption{Experimental specifications for Planck \cite{:2006uk}, SPIDER \cite{Fraisse:2011xz} and CMBPol \cite{Bock:2009xw}. For each experiment, we list the observed fraction $f_{sky}$ of the sky, the channel frequency in GHz, the FWHM in arcminutes, the sensitivity per pixel for the Stokes $I$ ($\sigma_T$), $Q$ and $U$ ($\sigma_P$) parameters in $\mu$K.}
\label{tabexp}
\end{center}
\end{table}
We consider a detector noise of $(\theta\sigma)^2$ for each frequency channel 
where $\theta$ is the FWHM of the beam assuming a Gaussian profile and
$\sigma$ is the sensitivity.
We add to each fiducial spectrum $C_\ell$, calculated with CAMB \cite{Lewis:1999bs}, a noise
spectrum given by
\begin{equation}
N_\ell = (\theta\sigma)^2\,e^{l(l+1)/l_b^2} \, ,
\end{equation}
where $l_b \equiv \sqrt{8\ln2}/\theta$.
In the analysis, we assume that beam and foreground uncertainties are smaller than the statistical errors.

The Fisher matrix is defined as
\begin{equation}
F_{ij}\equiv \Bigl\langle -\frac{\partial^2 
\ln \mathcal{L}}{\partial p_i \partial p_j}\Bigr\rangle _{p_0} \, ,
\end{equation}
where $\mathcal{L}({\rm data}|{\bf{p}})$ is the likelihood function of a set
of parameters ${\bf p}$ given some data; the partial derivatives and the
averaging are evaluated using the fiducial values ${\bf p_{0}}$ of the
parameters.  The Cram\'er-Rao inequality implies that $(F^{-1})_{ii}$ is the
smallest variance in the parameter $p_i$, so 
we can generally think of $F^{-1}$ as the best possible covariance
matrix for estimates of the vector ${\bf p}$. The 1-$\sigma$ error for each
parameter is then 
\begin{equation}\label{sigma}
\sigma_{p_{i}} = \sqrt{(F^{-1})_{i i}}\,.
\end{equation}

The Fisher matrix for a CMB experiment is (see \cite{Bond:1997wr})
\begin{equation}
     F^{\rm CMB}_{i j} = \sum_{l=2}^{l_{\rm max}} 
     \frac{\partial C_l}{\partial p_i}
     ({\rm Cov}_l)^{-1}
     \frac{\partial C_l}{\partial p_i},
\label{fishercmb}
\end{equation}
where ${\rm Cov}_l$ is the spectra covariance matrix.  We use
information in the power spectra up to $l_{\rm max}=2500$. The partial derivative $\frac{\partial C_l}{\partial \gamma}$ is analytical in $\gamma=0$:
\begin{equation}
 \frac{\partial C_l}{\partial \gamma} \equiv (-2 \gamma)C_\ell^{ad} + 2 \gamma C_\ell^{nid} 
 -\frac {2 (1-2\gamma^{2})}{\sqrt{(1-\gamma^{2})}}C_\ell^{corr} \, .
\end{equation}
As anticipated above, the parameterization in terms of $\gamma$, instead of $\alpha=\gamma^2$ as often seen in the literature, 
cancels the divergence of the partial derivative $\frac{\partial C_l}{\partial \alpha}$ in $\alpha=0$. 
Thus this parameterization allows us to use the Fisher matrix formalism for the fiducial value $\gamma=0$.

In the present analysis, we take as a fiducial model a flat $\Lambda$CDM model 
with parameter values given by the WMAP7 measurements\footnote{\url{http://lambda.gsfc.nasa.gov/product/map/current/params/lcdm_sz_lens_wmap7.cfm}}, i.e. $\Omega_b h^2=0.02258$ and $\Omega_{dm} h^2=0.1109$, the optical depth
to reionization $\tau=0.088$, $H_0=71\, {\rm km/s/Mpc}$, the spectral index $n_s=0.963$,
and the amplitude of the curvature perturbation $\Delta^2_{\mathcal R}(k_0) = 2.43\times 10^{-9}$. 
We consider three families of massless neutrinos, but we checked that taking massive neutrinos with total mass $M_\nu = 0.6 \eV$ neutrinos did not affect the results. 
Finally, we take the fiducial values $\DNeff=0$, $\gamma = 0$.

\begin{figure}[htb!]
\includegraphics[width=\columnwidth]{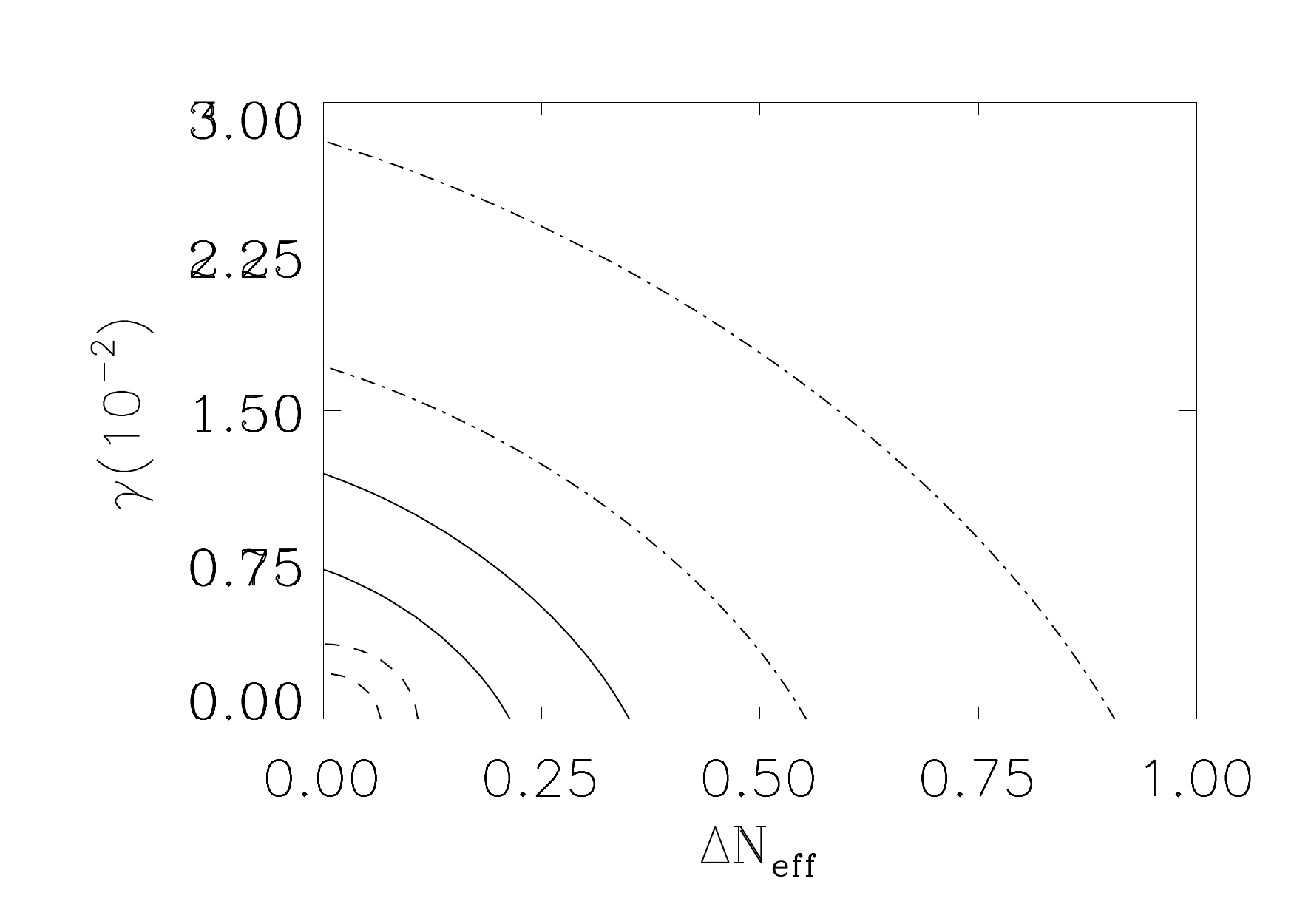} 
\caption{68\% and 95\% c.l.\ likelihood contours for Planck (solid line), SPIDER (dot-dashed line)
CMBPol (dashed line).} 
\label{gamma}
\end{figure}

The results of our analysis are shown in Figure~\ref{gamma}, where we draw the 2-dimensional likelihood in the
$\DNeff$-$\gamma$ plane for Planck, SPIDER and CMBPol. The corresponding
$1$-$\sigma$ constraints for $\gamma$ and $\DNeff$ are reported in Tab.~\ref{res}.
\begin{table}[!htb]
\begin{center}
\begin{tabular}{rccccc}
& fiducial value & $\sigma$(Planck) & $\sigma$(SPIDER) & $\sigma$(CMBPol)\\
\hline
& & & & \\
$\gamma$ & 0.0 & $5.3 \cdot 10^{-3}$ & $1.2\cdot10^{-2}$ & $1.5 \cdot 10^{-3}$\\
$\DNeff$ & 0.0 & 0.16 & 0.40 & 0.043\\
\end{tabular}
\caption{$1$-$\sigma$ constraints for $\gamma$ and $\DNeff$, for the Planck, SPIDER and CMBPol experiments.}
\label{res}
\end{center}
\end{table}
\section{Comparison with BBN constraints} \label{BBN}
Big Bang nucleosynthesis, and in particular  the primordial helium abundance $Y_p$, is recognized to be the most sensitive cosmic ``leptometer''  presently available, see for example~\cite{Serpico:2005bc,Simha:2008mt} or the review~\cite{Iocco:2008va}. So, it would be interesting to compare BBN constraints to the ones derived above. This task is made non-trivial by the fact that BBN is sensitive to different parameters than the CMB in particular, to a combination of the role of $N_{\rm eff}$, entering the expansion rate of the universe, and in principle to all the parameters describing the distribution of the $\nu_e$-flavour  neutrinos. In the case of interest, which assumes flavour-independent parameters and gaussian distributions, the only two independent parameters turn to be  $\bar{\xi}$ and $\sigma_\xi$, with $\Delta N_{\rm eff}$ fully specified in terms of them, but subleading and essentially negligible for the values $\bar{\xi}\ll 1$ of interest here.
Even assuming that the average value $\bar\xi$ is scale-independent, a slight dependence on the scale is expected for the width of the distribution of fluctuations.
Let us fix (arbitrarily) $\sigma_\xi$ at a scale $\lambda_{\rm BBN}$,  roughly corresponding to the horizon size at the time of BBN, of the order of $\sim {\cal O}(100)$ comoving parsecs. Namely, we fix $\sigma^2_\xi \sim \Delta^2_\xi (k_{\rm BBN})$ where $k_{\rm BBN} = 2\pi/\lambda_{\rm BBN}\equiv 6\times 10^{4}\,\mathrm{Mpc}^{-1}$. The CMB constraints can be translated into $\sigma^2_\xi$  by just  evaluating $\Delta^2_\zeta(k_{\rm BBN})$ (given that $\Delta^2_\iso$ has the same scale-dependence). Using WMAP7 best fit values $\Delta^2_\zeta (k=0.002\,\mathrm{Mpc}^{-1}) = 2.42\times 10^{-9}$ and $n_s = 0.966$ gives $\Delta^2_\zeta (k_{\rm BBN}) = 1.35\times 10^{-9}$. A first important consequence of this estimate is that the order of magnitude of the present constraints from CMB on $\alpha$ also holds for BBN-relevant fluctuations. In turn, it can be seen that this implies that $\sigma_\xi$ is very small.
This is an important information, since it allows us to use the predictions of {\it homogeneous}, degenerate BBN to infer the results of an otherwise 
{\it inhomogeneous} degenerate BBN scenario (see~\cite{Stirling:2002bj} for an early study of this subject). 
In fact, for a gaussian probability distribution for $\xi$,
\begin{equation}
P(\xi)=(2\pi\sigma_\xi^2)^{-1/2}\exp\left[-(\xi-\bar{\xi})^2/(2\,\sigma_\xi^2)\right]\,,
\end{equation}
one can estimate, for a generic nuclide abundance $X$,
\begin{eqnarray}
\langle X\rangle&=&\int P(\xi)  [X(\bar{\xi})+X'(\bar{\xi})(\xi-\bar{\xi})+{\cal O}((\xi-\bar{\xi})^2)]d\xi\nonumber\\
&=&X(\bar{\xi})+{\cal O}(\sigma_\xi^2)\,.
\end{eqnarray}
The vanishing of the integrand linear in $\xi$ depends on the fact that $P(\xi)$ is an even 
function of $\xi-\bar{\xi}$. Additionally, if the curvature of the function $X(\xi)$ is relatively small (as it happens to be, see Fig. 13 in~\cite{Iocco:2008va}) the approximation $\langle X\rangle\approx X(\bar{\xi})$ works even better (see also figs. in~\cite{Stirling:2002bj}). We estimated that even for a value as large as $\sigma_\xi\simeq 0.1$  the error of the approximation with respect to a proper averaging is of $\sim 0.6\%$ for deuterium  (hence well below the observational error) or of the order of $0.3\%$ for helium-4, comparable with the theoretical error and well below the error on the observations. For smaller $\sigma_\xi$, it scales as $\sigma_\xi^2$  and becomes soon negligible. As a consequence, the bounds computed in {\it homogeneous}, degenerate BBN can be used, to an excellent approximation, also for the case at hand. Needless to say, this also implies that BBN may give excellent constraints on $\bar\xi$, but it is insensitive to physically relevant values of the fluctuation $\sigma_\xi$.

By using the same conservative input as in~\cite{Mangano:2011ar} (fourth line in their Table I), we obtain the bounds 
\begin{equation}
{\bar\xi}^{\rm min}=-0.055\,,\:\:\:\bar{\xi}^{\rm max}=0.12\,,\label{xiBBNc}
\end{equation}
corresponding to the value below/above which only 5\% of the area of the marginalized distribution of probability of $\xi$ lies, respectively.
The BBN computation is based on the PArthENoPE code \cite{Pisanti:2007hk}. 
 
In order to illustrate the synergy between BBN and CMB, it turns useful to translate the CMB forecasts in the $\bar{\xi}-\sigma_\xi$ plane. For simplicity, let us write down the relation between variables in the (plausible) assumptions of  $\delta \xi \ll 1 $ and $\delta \xi \ll \bar\xi$. Then,  the relation between the power spectrum of the isocurvature perturbation $\iso_\nu$ to that of $\xi$ writes
\begin{equation}
\iso_\nu =\frac{3}{\pi} \frac{\frac{\bar \xi}{\pi} + \frac{\bar \xi^3}{\pi^3}}{\frac{7}{15} + \frac{2\bar \xi^2}{\pi^2}+\frac{\bar \xi^4}{\pi^4}}\delta \xi\equiv F(\bar\xi) \delta \xi \,.
\end{equation}

The relation above implies $\Delta^2_\iso = F(\bar{\xi})^2 \Delta^2_\xi$, so that
\begin{equation}
\Delta^2_\iso(k_{\rm BBN}) = F(\bar{\xi})^2 \sigma^2_\xi \, \Rightarrow \alpha \simeq 7.4\times 10^8 F(\bar\xi)^2 \sigma^2_\xi \, ,
\end{equation}
where we have used the fact that the data constrain $\alpha$ to be ${\mathcal O}(0.01)$ or less. Recalling that $\gamma=\sqrt{\alpha}$, we finally get the relation that we were looking for
\begin{equation}
\gamma \simeq 2.7\times 10^4 F(\bar\xi) \sigma_\xi \, . 
\end{equation}
\begin{figure}[htb!]
\includegraphics[width=\columnwidth]{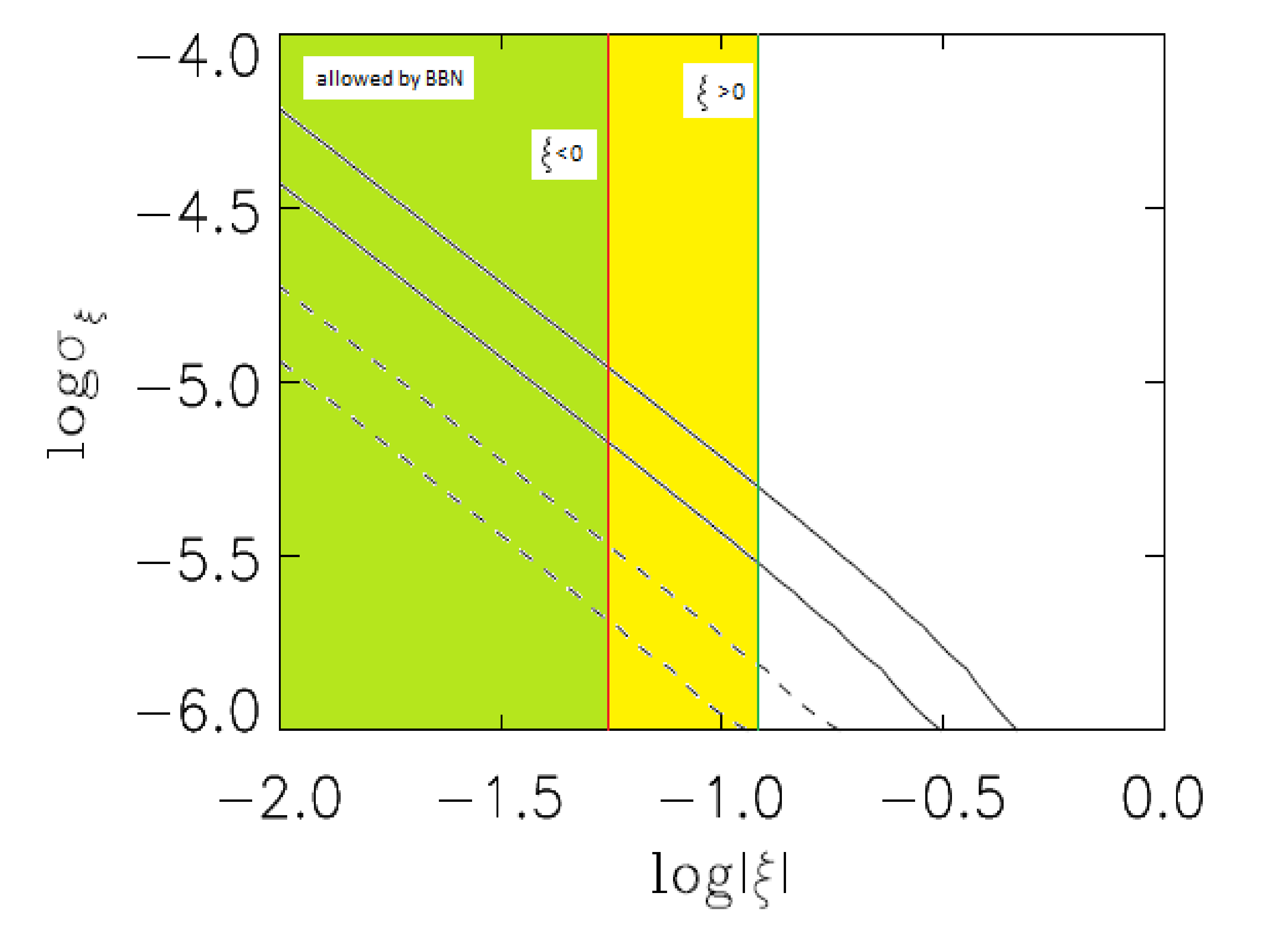} 
\caption{68\% and 95\% c.l.\ likelihood contours in the $(\log|\bar\xi|,\,\log\sigma_\xi)$ plane for Planck (solid line) 
and CMBPol (dashed line). The BBN allowed region are also shown (left of vertical lines), corresponding to the case considered in Eq.~(\ref{xiBBNc}).} 
\label{xi}
\end{figure}

On the other hand, recalling that $\Neff = \sum A_i$, with $A_i$ given by Eq. \reff{eq:Ai}, we can translate the bounds from the $(\gamma,\,\DNeff)$ plane to the $(\bar \xi,\, \sigma_\xi)$ plane. For this analysis, we only consider Planck and CMBPol since they give the better constraints on the parameters and can possibly become competitive with BBN in this respect. The two-dimensional 68\% and 95\% confidence regions for Planck and CMBPol are shown in Fig. ~\ref{xi}, along with the present BBN constraints on $\bar \xi$ reported in Eq.~(\ref{xiBBNc}).
\section{Conclusions}\label{Conclusions}

We have considered models in which a residual component of isocurvature fluctuations, and consequently a non-zero chemical potential, are generated in the neutrino component after curvaton decay. Using Fisher matrix techniques, we have assessed the constraints that Planck, SPIDER and CMBPol will be able to put on the amplitude of the isocurvature component and on the extra energy density associated to the non-vanishing neutrino chemical potential. These bounds have been expressed in terms of the average value of the degeneracy parameter $\bar \xi$ and its spatial variance $\sigma^2_\xi$, and then compared with the constraints resulting from the effect of neutrino degeneracy on BBN. While the latter is only sensitive to the mean value $\bar \xi$, for small $\sigma_\xi$,  CMB data provide a negative correlation between $\bar \xi$ and $\sigma_\xi$:
for $\delta \xi \ll \bar \xi $ (on which our analysis is based), large values of the fluctuations are allowed for sufficiently small $\bar \xi$.

In particular, assuming a fiducial model with purely adiabatic primordial fluctuations, we find that the future experiments will sensibly improve the constraints on the non-adiabaticity parameter $\gamma$ and on the effective number of neutrino families $\DNeff$. The current 95\% C.L. WMAP bound corresponds to $\gamma\lesssim 0.1$; we find that, at the same level, SPIDER will be able to constrain $\gamma$ below $2.4\cdot 10^{-2}$, representing an improvement of a factor 4. Planck and CMBPol will be able to put 95\% C.L. upper limits $\gamma < 1.1\cdot 10^{-2}$ and $\gamma < 3.0\cdot 10^{-3}$, i.e. to improve by a factor 10 and 30 over current data, respectively. For what concerns $\DNeff$, we already noticed how WMAP only provides a lower bound for this quantity. In general, current data allows to constrain this parameter with precision $\sigma(\DNeff)\simeq 1$ or smaller (see e.g. Ref. \cite{Hamann:2007pi} for a detailed analysis). In terms of the average value of the degeneracy parameter, this reads\footnote{We note that for finite neutrino masses the effects of $\bar\xi$ on the CMB observables are not completely encoded by $\DNeff$ \cite{Lesgourgues:1999wu,Lattanzi:2005qq}. Joint constraints on $\bar \xi$ and $\DNeff$ have been derived for example in Refs. \cite{Lattanzi:2005qq,Popa:2008tb}.} $\sigma(\bar\xi) \simeq 0.9$, which is quite large with respect to the BBN bound. For comparison, we find that if $\DNeff=0$, Planck, Spider and CMBPol will be able to bound $\DNeff \lesssim 0.3,\, 0.8,\, 0.08$ at the 95\% C.L., respectively, corresponding to $\bar\xi < 0.5, \,0.8,\, 0.24$. Although these values represent considerable improvements over present CMB constraints, they show that future CMB experiments, with the partial exception of CMBPol, will still be unable to compete with BBN in this respect. Moreover BBN is the only one sensitive to \textit{the sign} of $\bar\xi$.
Both effects are actually due to the dominant weak interaction probe provided by BBN,
as opposed to purely gravitational effect to which CMB is sensitive.

\section*{Acknowledgments}
The authors acknowledge useful correspondence on the curvaton scenario with David Wands. 
The authors would like to thank Matteo Martinelli for useful discussion and help.


\end{document}